\documentclass[%
unsortedaddress,
nofootinbib,
 amsmath,amssymb,
 aps,
twocolumn,
]{revtex4}

\usepackage{graphicx}
\usepackage{dcolumn}
\usepackage{bm}

\usepackage{braket}
\usepackage{xcolor}
\usepackage{cancel}
\usepackage{titlesec}
\usepackage{etoolbox}
\usepackage{subfig}
\usepackage{mathtools}
\usepackage[normalem]{ulem}
\usepackage[all]{nowidow}

\newcommand{\be}{\begin{equation}}
\newcommand{\ee}{\end{equation}}
\newcommand{\kb}{k_{\mathrm B}}
\newcommand{\dd}{\!\mathrm{d}}

\newcommand{\um}[1]{\ensuremath{\mathrm{\,#1}}}

\newcommand{\inv}{^{-1}}

\newcommand{\stkout}[1]{\ifmmode\text{\sout{\ensuremath{#1}}}\else\sout{#1}\fi}

\begin{document}

\preprint{APS/123-QED}

\title{Learning Mappings between Equilibrium States of Liquid Systems \\ Using Normalizing Flows}

\author{Alessandro Coretti 
}
\email{alessandro.coretti@univie.ac.at}
\affiliation{University of Vienna, Faculty of Physics, 1090 Vienna, Austria.}

\author{Sebastian Falkner 
}
\affiliation{University of Vienna, Faculty of Physics \& Vienna Doctoral School in Physics, 1090 Vienna, Austria}

\author{Phillip Geissler 
}
\thanks{Phillip L. Geissler passed away on July 17, 2022, while this paper was being prepared for submission.}
\affiliation{Department of Chemistry, University of California, Berkeley, California 94720, USA}

\author{Christoph Dellago 
}
\affiliation{University of Vienna, Faculty of Physics, 1090 Vienna, Austria.}
%


\date{\today}

\begin{abstract}
Generative models are a promising tool to address the sampling problem in multi-body and condensed-matter systems in the framework of statistical mechanics. In this work, we show that normalizing flows can be used to learn a transformation to map different liquid systems into each other allowing at the same time to obtain an unbiased equilibrium distribution through a reweighting process. Two proof-of-principles calculations are presented for the transformation between Lennard-Jones systems of particles with different depths of the potential well and for the transformation between a Lennard-Jones and a system of repulsive particles. In both numerical experiments, systems are in the liquid state. In future applications, this approach could lead to efficient methods to simulate liquid systems at \emph{ab-initio} accuracy with the computational cost of less accurate models, such as force field or coarse-grained simulations.
\end{abstract}

\maketitle

The generation of properly weighted samples from physical distributions is a central problem in statistical mechanics and computational physics. Monte Carlo and molecular dynamics simulations are the state of the art approaches for producing samples from equilibrium distributions, but they do not come without drawbacks. Being both based on sequential updates of configurations, they can be slow in exploring configuration space due to statistical correlations between subsequent configurations. As a result, it is computationally expensive to generate completely decorrelated samples. In complex systems with rare transitions, these correlations can become a limiting factor of the sampling efficiency.

Machine learning and, in particular, generative models~\cite{goodfellow_advances_2014,kingma__2014,papamakarios_normalizing_2021} are a promising alternative to overcome some of these problems. These models are able, in principle, to generate decorrelated samples from an arbitrary distribution at the cost of having a relatively large pool of examples from which the model can be trained. In the case of statistical physics systems, this may appear as circular reasoning as many (independent) samples are needed to produce many (independent) samples. Nonetheless, it has been shown that the knowledge of the analytical form of the target distribution can be used either for enhancing the training or to compensate for a biased generated distribution resulting from the limited expressiveness of the network. These matters have been addressed by No\'e and coworkers~\cite{noe_boltzmann_2019}, who have used, for the first time, normalizing flows --- a particular class of generative models based on deep neural networks --- to sample equilibrium configurations of statistical mechanics systems. The principle at the basis of these models, also called Boltzmann generators, is to train a network to transform samples from an elementary, easy-to-sample, ``prior'' distribution, such as a multivariate Gaussian or uniform distribution, into a more complex and physically interesting distribution such as the Boltzmann distribution.

Another line of research with a similar approach was pursued by Wirnsberger and coworkers~\cite{wirnsberger_normalizing_2021}, in which, again, a normalizing flow is trained to transform easy-to-sample distributions into more complex and physically meaningful ones. The main difference with the approach by No\'e et al. is the introduction of a prior that is closer to the physical distribution to be generated.
For example in Ref.~\citenum{wirnsberger_normalizing_2021}, following a line of reasoning already sketched in a previous publication~\cite{wirnsberger_targeted_2020}, a network is trained to transform a superposition of Gaussians placed at lattice sites of a particular crystal structure into a physical distribution of a crystalline solid. The encoding of physical information in the prior allows the training of the network to be performed without samples from the target distribution using only its potential energy function.

In this letter we combine the ideas introduced in these two lines of research and we show that an invertible transformation can be learned by a neural network to transform one Boltzmann distribution into another, different, Boltzmann distribution and that the generated configurations can be successfully reweighted to obtain unbiased samples. The approach is successfully tested on liquids at different thermodynamic points in the phase diagram, showing that this Boltzmann-to-Boltzmann approach can be applied to disordered phases of a system. 

We consider two systems A and B, whose configurations $x_{\mathrm{A}}$ and $x_{\mathrm{B}}$ are distributed according to the probability densities $\rho_{\mathrm{A}}(x_{\mathrm{A}})$ and $\rho_{\mathrm{B}}(x_{\mathrm{B}})$. We define an invertible mapping $\Phi$ that transforms configurations of one system into the ones of the other and vice versa, i.e. $x_{\mathrm{B}} = \Phi(x_{\mathrm{A}})$ and $x_{\mathrm{A}} = \Phi\inv(x_{\mathrm{B}})$. The direction of the map is arbitrary and our choice is done without loss of generality. Under the application of $\Phi$, the probability distributions for systems A and B are related by 
\begin{subequations}
\label{eq:prob_map}
\begin{align}
    \rho_{\mathrm{B}}(x_\mathrm{B}) &= \rho_{\mathrm{A}}(x_\mathrm{A})\lvert\det \mathcal{J}(x_\mathrm{A})\rvert\inv , \label{eq:prob_map_a}\\
    \rho_{\mathrm{A}}(x_\mathrm{A}) &= \rho_{\mathrm{B}}(x_\mathrm{B})\lvert\det \mathcal{J}\inv(x_\mathrm{B})\rvert\inv , \label{eq:prob_map_b}
\end{align}
\end{subequations}
where $\mathcal{J}$ is the Jacobian matrix of $\Phi$ and $\mathcal{J}\inv$ is that of the inverse mapping, so that we have $\mathcal{J}\inv(x) \equiv \mathcal{J}\bigl(\Phi\inv(x)\bigr)$.

The analytical form of the transformation $\Phi$ is in general very hard if not impossible to obtain. On the other hand, an approximation $F$ can be found by means of a machine-learning approach.
A class of deep neural networks well suited to this task are the so-called normalizing flows~\cite{rezende_variational_2015}, a particular kind of generative models characterized by the invertibility of the transformation and by the ability of transforming probability distributions.
The application of the machine-learned transformation $F$ and it inverse $F\inv$ to samples from $\mathrm{A}$ and $\mathrm{B}$ gives the relations
\begin{subequations}
\label{eq:prob_map_ML}
\begin{align}
    P_{\mathrm{B}}(x_\mathrm{B}) &= \rho_{\mathrm{A}}(x_\mathrm{A})\lvert\det J(x_\mathrm{A})\rvert\inv , \label{eq:prob_map_ML_a}\\
    P_{\mathrm{A}}(x_\mathrm{A}) &= \rho_{\mathrm{B}}(x_\mathrm{B})\lvert\det J\inv(x_\mathrm{B})\rvert\inv , \label{eq:prob_map_ML_b}
\end{align}
\end{subequations}
where $P_{\mathrm{B}}$ and $P_{\mathrm{A}}$ are the generated probability distributions --- different, in general from the \emph{true} probability distributions $\rho_{\mathrm{B}}$ and $\rho_{\mathrm{A}}$ --- and where $J$ and $J\inv$ are the Jacobians of $F$ and $F\inv$, respectively.\footnote{The mapping $F$ and its Jacobian $J$ --- together with their inverse --- are also functions of the network parameters $\theta$ as in, for example, $F(x)\equiv F(x; \theta)$. We omit this dependence in the rest of this letter to keep the notation as light as possible.}

Equations~\eqref{eq:prob_map_ML} enable the definition of a ``distance'' between the target and the generated distributions from which a loss function for the training process can be derived. The Kullback-Leibler (KL) divergence acts as such a statistical ``distance'':  Given two distributions $f(x)$ and $g(x)$ their KL divergence is given by
$$
\mathrm{KL}(g \parallel f) = \int \dd x g(x)\bigl[\log g(x) - \log f(x)\bigr] .
$$

We consider at first the direct transformation $F$ so that samples are transformed from the space $\mathrm{A}$ to the space $\mathrm{B}$. 
The KL divergence between the target and the generated distribution is given by
\begin{equation}
    \label{eq:KL_B}
    \mathrm{KL}(\rho_B \parallel P_B) = \int \dd x_B \rho_B(x_B)\bigl[\log \rho_B(x_B) - \log P_B(x_B)\bigr] .
\end{equation}
Using Eqs.~\eqref{eq:prob_map_ML} we know how the map $F$ acts on the original samples from the source distribution $\rho_A$ and we can use this information to write Eq.~\eqref{eq:KL_B} as
$$
\begin{aligned}
    &\mathrm{KL}(\rho_B \parallel P_B) = \int \dd x_B \rho_B(x_B)\log \rho_B(x_B) + \\
    &- \int \dd x_B \rho_B(x_B)\log\bigl[
    \rho_{\mathrm{A}}\bigl(F\inv(x_{\mathrm{B}})\bigr)\lvert\det J\inv(x_{\mathrm{B}})\rvert\bigl] ,
\end{aligned}
$$
where we used the property that $\lvert\det J(x_\mathrm{A})\rvert\inv = \lvert\det J\inv(x_\mathrm{B})\rvert$.
The first term of the RHS is the negative entropy of the system $\mathrm{B}$ and does not depend on the transformation $F$ (nor on its inverse). Therefore, for the purpose of the definition of a training loss function, it is irrelevant. The second term of the RHS can be recast into the form of an average over the configurations of the system B. Therefore, we can define the loss function for the direct transformation as
\begin{equation}
    \label{eq:L_inv}
    L_{\text{dir}} = -\Braket{\log\rho_{\mathrm{A}}\bigl(F\inv(x_{\mathrm{B}})\bigr)+\log\lvert\det J\inv(x_{\mathrm{B}})\rvert}_{\mathrm{B}} .
\end{equation}
Applying the same line of reasoning for the inverse transformation, i.e. starting from $\mathrm{KL}(\rho_A \parallel P_A)$, we obtain the loss function
\begin{equation}
    \label{eq:L_dir}
    L_{\text{inv}} = -\Braket{\log\rho_{\mathrm{B}}\bigl(F(x_{\mathrm{A}})\bigr)+\log\lvert\det J(x_{\mathrm{A}})\rvert}_{\mathrm{A}} .
\end{equation}
The total loss function for the training of the network is  $L=\lambda_{\text{dir}}L_{\text{dir}}+\lambda_{\text{inv}}L_{\text{inv}}$, where $\lambda_{\text{dir}}$ and $\lambda_{\text{inv}}$ are tunable parameters used to focus the training on the direct transformation or on its inverse.

The definition of the loss function through Eqs.~\eqref{eq:L_inv} and~\eqref{eq:L_dir} highlights the crucial role that the Jacobian of the transformation $J$ has in the training process. An efficient implementation of an ML solution for the problem discussed in this letter inevitably involves an efficient implementation of the computation of $\log \lvert\det J(x)\rvert$ and its inverse.

This goal can be achieved with neural networks based on a split-coupling flow architecture~\cite{dinh_nice_2015, dinh_density_2017}. The idea at the basis of these networks is to split the input vector $x$ in two \emph{channels} $x^{\mathrm{I}}$ and $x^{\mathrm{II}}$ and to perform invertible operations only on one of the channels, in such a way to obtain a transformation matrix which is triangular and whose determinant is easy and fast to compute.
Moreover, as invertibility of a map is closed under function composition, many such maps operating on the two channels of the input alternately can be stacked together, so that the whole input vector gets transformed and the expressiveness of the network is improved.
An example of these transformations are the so-called real non-volume preserving (real-NVP) networks~\cite{dinh_density_2017} in which only simple operations of scaling and translation are performed on one channel of the input. Assuming a splitting of the input $x = (x^{\mathrm{I}},x^{\mathrm{II}})$, a real-NVP \emph{block} acts so to obtain the output $y=(y^{\mathrm{I}},y^{\mathrm{II}})$ as
\begin{equation}
\label{eq:realnvp_block}
\begin{aligned}
    y^{\mathrm{I}} &= x^{\mathrm{I}} ,\\
    y^{\mathrm{II}} &= x^{\mathrm{II}} \odot \exp\bigl[S(x^{\mathrm{I}})\bigr] + T(x^{\mathrm{I}}) ,
\end{aligned}
\end{equation}
where $S,T: \mathbb{R}^{\dim x^{\mathrm{I}}} \to \mathbb{R}^{\dim x^{\mathrm{II}}}$ are obtained via arbitrary, trainable, nonlinear transformations and the exponential and the multiplication (represented by the symbol $\odot$) are applied element-wise. Denoting the Jacobian matrix of the transformation in Eq.~\eqref{eq:realnvp_block} by $j$, its log-determinant can easily be computed as
$$
    \log\lvert\det j (x)\rvert = \sum_{\alpha}S_{\alpha}(x^{\mathrm{I}}) ,
$$
where $S_{\alpha}$ are the components of $S$.
Note that in general, the nonlinear transformations $S$ and $T$ are not invertible, but the transformation in Eq.~\eqref{eq:realnvp_block} is, with inverse
$$
\begin{aligned}
    x^{\mathrm{I}} &= y^{\mathrm{I}} ,\\
    x^{\mathrm{II}} &= \bigl[y^{\mathrm{II}} - T(y^{\mathrm{I}})\bigr] \odot \exp\bigl[-S(y^{\mathrm{I}})\bigr] ,
\end{aligned}
$$
and log-determinant
$$
    \log\lvert\det j\inv (y)\rvert = -\sum_{\alpha}S_{\alpha}(y^{\mathrm{I}}) .
$$
As real-NVP blocks are stacked together one after the other, the final transformation $F$ results from a composition of single transformations like in Eq.~\eqref{eq:realnvp_block}. Consequently the determinant can be obtained as the product of determinants of the single blocks, i.e. $\lvert\det J(x)\rvert = \prod_{k}\lvert\det j_k(x)\rvert$ where the product runs over all real-NVP blocks.

We here focus the derivation on molecular systems in the canonical ensemble for which the ground truth distributions $\rho_A$ and $\rho_B$ have the form of the Boltzmann distribution
$$
    \rho(x) \propto \exp[-\beta U(x)] ,
$$
where $U$ is the potential energy function of the particular system and $\beta$ is the inverse temperature $\beta\inv = \kb T$.
It is then interesting to look back at the loss functions derived in Eqs.~\eqref{eq:L_inv} and~\eqref{eq:L_dir}, which, after substitution of the particular form of $\rho$, become
\begin{align*}
        L_{\text{dir}} &= \Braket{\beta_{\mathrm{A}} U_{\mathrm{A}}\bigl(F\inv(x_{\mathrm{B}})\bigr)-\log\lvert\det J\inv(x_{\mathrm{B}})\rvert}_{\mathrm{B}} ,\\
    L_{\text{inv}} &= \Braket{\beta_{\mathrm{B}} U_{\mathrm{B}}\bigl(F(x_{\mathrm{A}})\bigr)-\log\lvert\det J(x_{\mathrm{A}})\rvert}_{\mathrm{A}} ,
\end{align*}
where $U_{\mathrm{A}}$ and $U_{\mathrm{B}}$ are the potential energy functions of the systems A and B at inverse temperatures $\beta_{\mathrm{A}}$ and $\beta_{\mathrm{B}}$, respectively. The logarithms of the partition functions arising from $\log\rho_{\mathrm{A}}$ and $\log\rho_{\mathrm{B}}$ are constant with respect to the network parameters and are thus irrelevant for the definition of the training losses.

For the number of degrees of freedom that are relevant for statistical physics systems, even the most powerful network can be subject to a bias in the generated distribution, regardless of the length and of the accuracy of the training process. This bias results from the difference between the exact transformation $\Phi$ and the machine-learned one $F$. A systematic procedure to account for this difference becomes particularly important, for example, when no reference data are available and it is not possible to assess the accuracy of the results produced by the network. It is therefore crucial, if we aim to use the generated distributions to compute physical observables, to take into account the bias of the generated distribution and to compensate for it when computing averages. 
Indeed, given a set of $N$ configurations $\{x^i\}^N_{i=1}$ sampled according to the probability distribution $\rho(x)$, it is possible to approximate the expected value of an observable $\braket{O} = \int\dd x\, \rho(x) O(x)$ as
\begin{equation}
    \label{eq:observ}
    \braket{O} \approx \frac{1}{N}\sum_{i=1}^N O(x^i) ,
\end{equation}
where $O(x)$ is the microscopic estimator of the macroscopic observable.
However, Eq.~\eqref{eq:observ} does no longer yield the correct average if the configurations $\{x^i\}$ are sampled according to a probability $P(x)\neq\rho(x)$. In this case it is necessary to account for the difference in the relative weights of the sampled configurations. The average $\braket{O}$ can then be written as
\begin{equation}
    \label{eq:rew_observ}
    \braket{O} \approx \frac{\sum_{i=1}^N \omega(x^i)O(x^i)}{\sum^N_{i=1}\omega(x^i)} ,
\end{equation}
where $\omega(x^i)$ are the relative weights of the configurations $x^i$ given by~\cite{noe_boltzmann_2019}
\begin{equation}
    \label{eq:weights}
    \omega(x^i) = \frac{\rho(x^i)}{P(x^i)}.
\end{equation}
This expression for the weights can be verified by writing down the ensemble average for the microscopic estimator $O(x)$ computed under the distribution $\rho(x)$ and under the reweighted distribution $\omega(x)P(x)$:
$$
    \braket{O} = \int \dd x\, \rho(x)O(x) = \int \dd x\, \omega(x)P(x)O(x) ,
$$
which is true for $\omega(x)$ given by Eq.~\eqref{eq:weights}. Here, we assume $P(x^i) \neq 0$, which is guaranteed by the normalizing flow architecture, as samples of nonzero probability in the source space are mapped to points of nonzero probability in the target space.

Knowledge of how the probability of a sample transforms under the application of the machine-learned map $F$ makes it possible to give an explicit expression for the reweighting factors for the two systems A and B. For instance, considering Eq.~\eqref{eq:weights} for system $\mathrm{A}$ and substituting the expression of the generated probability $P_{\mathrm{A}}(x)$ as written in Eq.~\eqref{eq:prob_map_ML_b}, we get
$$
\omega_{\mathrm{A}}(x^i_\mathrm{A}) = \frac{\rho_{\mathrm{A}}(x^i_\mathrm{A})}{\rho_{\mathrm{B}}(x^i_\mathrm{B})\lvert\det J\inv(x^i_\mathrm{B})\rvert\inv} .
$$
Introducing now the explicit expressions for the equilibrium distributions $\rho_{\mathrm{A}}(x)$ and $\rho_{\mathrm{B}}(x)$ we obtain the reweighting factor for averages computed using configurations transformed from B to A as
\begin{equation}
\label{eq:omega_a}
\begin{aligned}
    \omega_{\mathrm{A}}(x^i_\mathrm{A}) &\propto \exp\biggl[-\beta U_{\mathrm{A}}(x^i_\mathrm{A}) +\\
    &+ \beta U_{\mathrm{B}}\bigl(F(x^i_\mathrm{A})\bigr) - \log\Bigl[\lvert\det J(x^i_{\mathrm{A}})\rvert\Bigr]\biggr] .
\end{aligned}
\end{equation}
The same kind of approach yields the reweighting factor for averages computed using configurations transformed from A to B as
\begin{equation}
\label{eq:omega_b}
\begin{aligned}
    \omega_{\mathrm{B}}(x^i_\mathrm{B}) &\propto \exp\biggl[-\beta U_{\mathrm{B}}(x^i_\mathrm{B}) + \\
    & +\beta U_{\mathrm{A}}\bigl(F\inv(x^i_\mathrm{B})\bigr) - \log\Bigl[\lvert\det J\inv(x^i_{\mathrm{B}})\rvert\Bigr]\biggr] .
\end{aligned}
\end{equation}
Note that the proportionality factors cancel out in Eq.~\eqref{eq:rew_observ}. The weights $\omega$ can be used either for standard reweighting of the observables computed as averages over the configurations or to directly resample the configurations by existing resampling methods, such as jackknifing or bootstrapping~\cite{efron_jackknife_1982}. Moreover, these weights have been shown to be connected to the free energy difference between the two systems and can be used for free energy estimation~\cite{wirnsberger_targeted_2020}. 

\begin{figure}[htbp]
\centering
\includegraphics[width=\columnwidth]{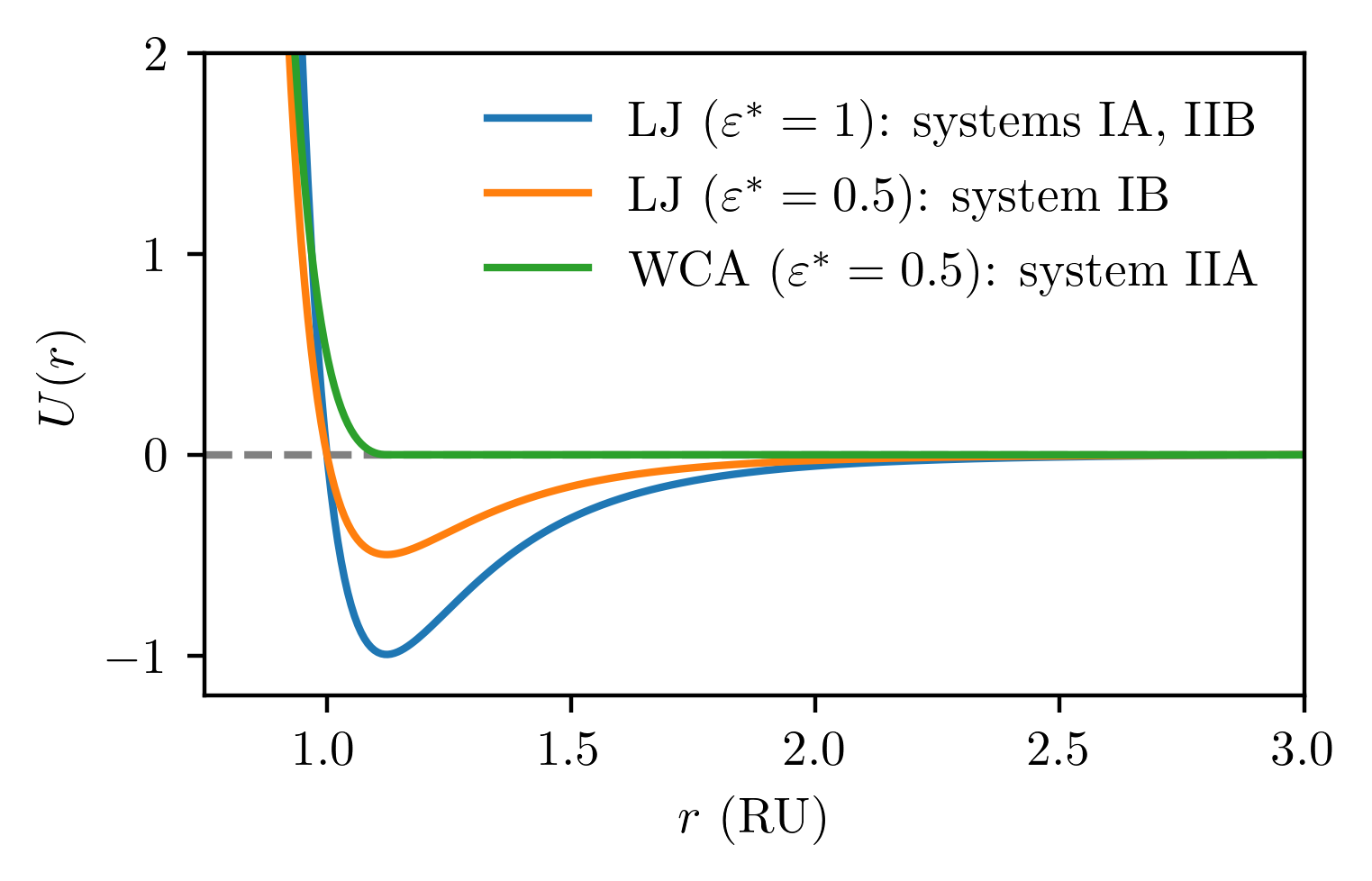}
\caption{Different pair potentials used for the numerical experiments in this work. The blue line represents the Lennard-Jones potential (Eq.\eqref{eq:LJ}) in standard reduced units (systems $\mathrm{IA}$ and $\mathrm{IIB}$), while the orange line represents the same potential when $\varepsilon$ is divided by a factor of two (system $\mathrm{IB}$). The green line is the WCA potential (Eq.~\eqref{eq:WCA}) with $\varepsilon_{\mathrm{IIA}} = \varepsilon_{\mathrm{IA}}/2$ (system $\mathrm{IIA}$).}
\label{fig:potential_energy}
\end{figure}

The approach presented here is tested and validated on a simple liquid system in two dimensions. 
In what follows, two examples are shown, where a normalizing flow is trained to map between two different systems in the liquid state.
In one case, we change the numerical value of a parameter of the interaction potential, while in the other, we also modify the functional form of the interaction. The different potential energy functions are shown in Fig.~\ref{fig:potential_energy}.

The detailed architecture of the networks together with the training protocols used to model the transformation $F$ in the two examples can be found in the SM. In the simulations, a training set of $9\times10^{3}$ configurations obtained via Markov chain Monte Carlo (MCMC) is prepared for the target system (see SM for more details). The network is validated using $10^{3}$ independent configurations. To ensure a good exploration of the source space during training, new configurations are continuously sampled from the source system using MCMC starting from a pool of independent configurations, which gets updated as training goes on. This is crucial to conserve memory of previous explored configurations, avoiding to get trapped in local minima of the source space. This \emph{on-the-fly} sampling also ensures that just the right amount of samples from the source space are produced during training.
After the training is complete, configurations are sampled from system $\mathrm{A}$ and system $\mathrm{B}$ and the learned transformation $F$ and its inverse $F\inv$ are applied, respectively. A reweighting procedure using the weights of Eq.~\eqref{eq:weights} is applied to the computed averages to obtain correctly weighted physical observables. Error bars are estimated via bootstrapping.

In the first calculation (I), a network is trained to transform between two systems of two-dimensional Lennard-Jones particles in the liquid state. Systems $\mathrm{IA}$ and $\mathrm{IB}$ are made up by $N=32$ particles interacting through the pair potential
\begin{equation}
\label{eq:LJ}
    U(r) = 4\varepsilon\biggl[\biggl(\frac{\sigma}{r}\biggr)^{12} - \biggl(\frac{\sigma}{r}\biggr)^6\biggr] ,
\end{equation}
where $r$ is the interparticle distance, see Fig.~\ref{fig:potential_energy}. Reduced units (indicated by the asterisk) are used throughout the paper, with $\sigma$ and $\varepsilon$ of system $\mathrm{IA}$ chosen as units of length and energy, respectively. All calculations are carried out at reduced temperature $T^* = 1$ and reduced density $\varrho^* = 0.735$. Periodic boundary conditions are enforced in all directions.
We set $\sigma_{\mathrm{IA}} = \sigma_{\mathrm{IB}}$, and the network is trained to map between $\varepsilon_{\mathrm{IA}}$ and $\varepsilon_{\mathrm{IB}}=\varepsilon_{\mathrm{IA}}/2$. These parameters are chosen so that the system states are well within the liquid part of the phase diagram~\cite{li_phase_2020}. Note that scaling $\varepsilon$ is equivalent to scaling $\beta$ by the same factor, i.e. dividing $\varepsilon$ by a factor of two means doubling the temperature of the system.

\begin{figure*}[htbp]
\centering
\includegraphics[width=.9\textwidth]{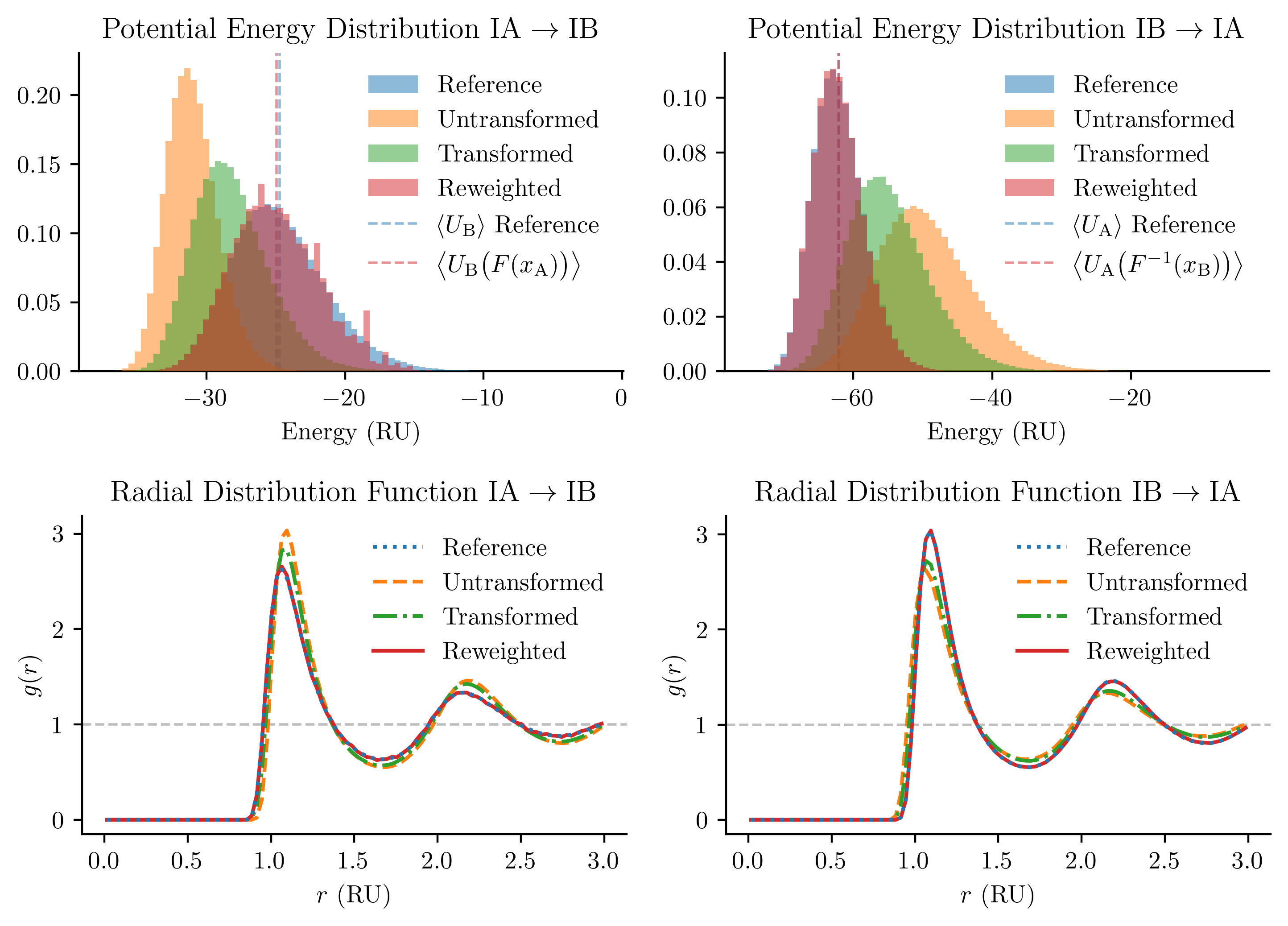}
\caption{Results obtained from a transformation connecting a standard Lennard-Jones system of particles in two dimensions at $T^*=1$ and $\varrho^*=0.735$ to the same system with $\varepsilon_{\mathrm{IB}} = \varepsilon_{\mathrm{IA}}/2$. Orange refers to the ``Untransformed'' configurations, i.e. the transformations applied without network training (identity mapping). Green refers to the raw result of the transformations after training. Red are the reweighted results. Blue are the observables computed using the configurations of the training set, taken as a reference. Top: Normalized histogram of potential energies of the configurations for the direct ($\mathrm{IA}\to\mathrm{IB}$, left) and inverse ($\mathrm{IB}\to\mathrm{IA}$, right) transformations. Error bars on the histogram bins would be nearly invisible and are therefore not shown. Dashed lines are the average potential energies computed using the reference (blue) and the reweighted (red) configurations. Bottom: Radial distribution functions for the direct ($\mathrm{IA}\to\mathrm{IB}$, left) and inverse ($\mathrm{IB}\to\mathrm{IA}$, right) transformation. Error bars are smaller than thickness of the drawn lines.}
\label{fig:epsilon}
\end{figure*}

The potential energy distributions and radial distribution functions of the system before and after the transformations in both directions ($\mathrm{IA}\to \mathrm{IB}$ and $\mathrm{IB}\to \mathrm{IA}$) are shown in Fig.~\ref{fig:epsilon} for a sample of $5\times 10^5$ configurations.
Additionally, in the same figure, the reweighted observables obtained as outlined above (Eqs.~\eqref{eq:omega_a} and~\eqref{eq:omega_b}) are shown. As reference we use the observables computed on the configurations of the training sets.
These results indicate that the network can be trained to transform the configurations sampled from the system with the potential $U_\mathrm{IA}$ into configurations typical for the system interacting via the potential $U_\mathrm{IB}$ and vice versa. 
We observe that it is harder for the network to transform configurations from a more ordered system into a more disordered one and to reweight them. This is apparent looking at the potential energy distribution of the samples transformed from system $\mathrm{IA}$ to system $\mathrm{IB}$ with $\varepsilon_{\mathrm{IB}} = \varepsilon_{\mathrm{IA}}/2$ (Fig.~\ref{fig:epsilon}), where high energy configurations are underrepresented. Nevertheless, the reweighting is successful in correcting for this bias and average values are in very good agreement. The signal to noise ratio of the weights grows with an increasing number of samples.

\begin{figure*}[htbp]
\centering
\includegraphics[width=.9\textwidth]{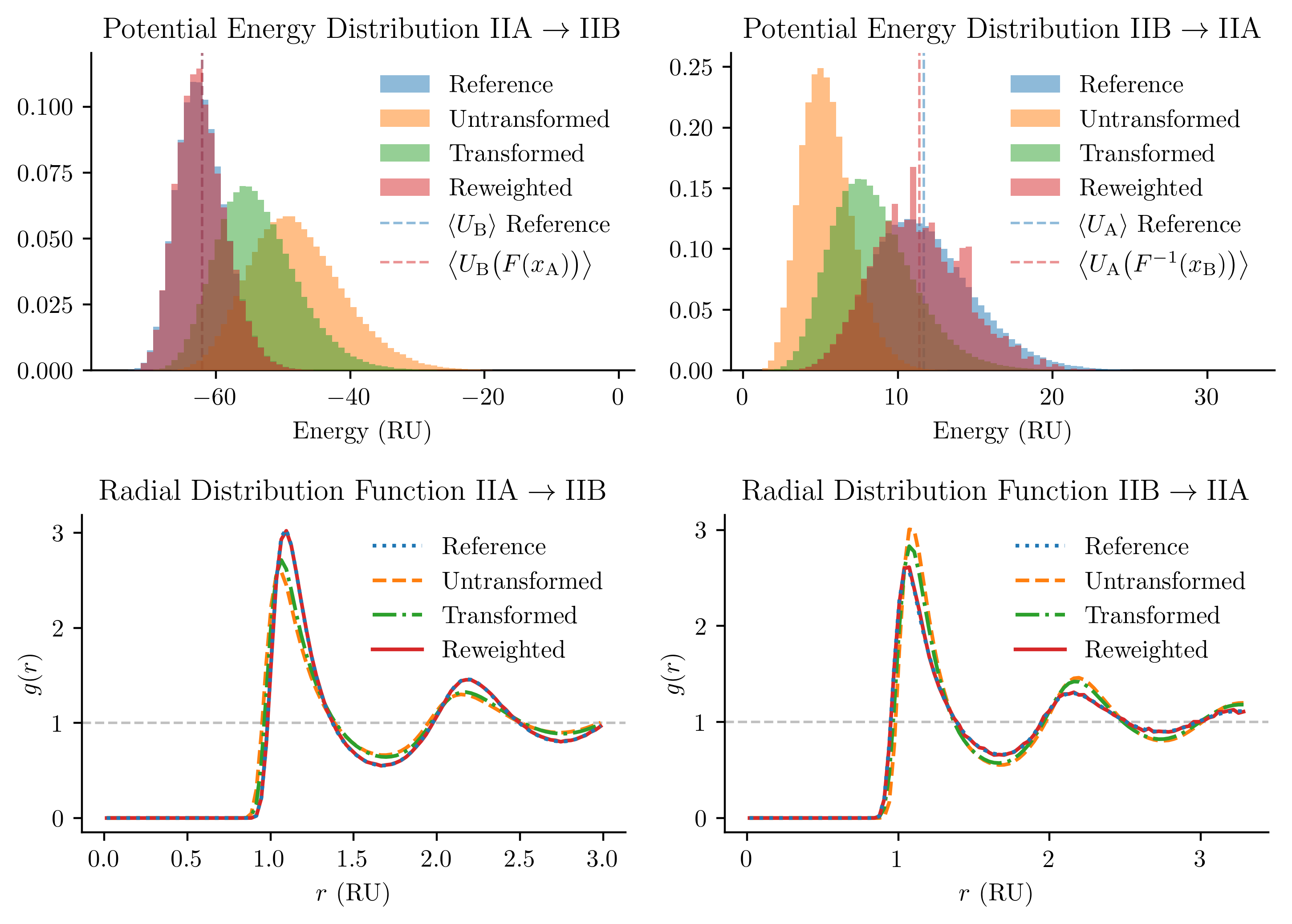}
\caption{Results obtained transforming a system of particles in two dimensions at $T^*=1$ and $\varrho^*=0.735$ interacting through a WCA potential ($\varepsilon_{\mathrm{IIA}} = \varepsilon_{\mathrm{IA}}/2$ and $\sigma_{\mathrm{IIA}} = \sigma_{\mathrm{IA}}$) into a system interacting via a Lennard-Jones potential ($\varepsilon_{\mathrm{IIB}} = \varepsilon_{\mathrm{IA}}$ and $\sigma_{\mathrm{IIB}} = \sigma_{\mathrm{IA}}$). Key is the same as Fig.~\ref{fig:epsilon}. Top: Normalized histogram of potential energies of the configurations for the direct ($\mathrm{IIA}\to\mathrm{IIB}$, left) and inverse ($\mathrm{IIB}\to\mathrm{IIA}$, right) transformations. Error bars on the histogram bins would be nearly invisible and are therefore not shown. Bottom: Radial distribution functions for the direct ($\mathrm{IIA}\to\mathrm{IIB}$, left) and inverse ($\mathrm{IIB}\to\mathrm{IIA}$, right) transformation. Error bars are smaller than the drawn lines.}
\label{fig:repulsive}
\end{figure*}
In the second example (II), we choose to map a system interacting through the Weeks-Chandler-Andersen (WCA) potential, whose form is defined as
\begin{equation}
\label{eq:WCA}
    U(r) = 
    \begin{dcases}
    4\varepsilon\biggl[\biggl(\frac{\sigma}{r}\biggr)^{12} - \biggl(\frac{\sigma}{r}\biggr)^6 + \frac{1}{4}\biggr] & \text{if} \ r < \sqrt[6]{2\sigma} ,\\
    0 & \text{otherwise} ,
    \end{dcases}
\end{equation}
into a standard Lennard-Jones system with a potential energy given by Eq.~\eqref{eq:LJ}, see again Fig.~\ref{fig:potential_energy}.
This choice for the source system amounts to considering only the repulsive part of the Lennard-Jones potential and ignoring the attractive part. The thermodynamic state of the two systems is the same of the previous example, i.e. $T^*=1$ and $\varrho^*=0.735$. The potential parameters are chosen to be $\varepsilon_{\mathrm{IIA}}=\varepsilon_{\mathrm{IA}}/2$ and $\sigma_{\mathrm{IIA}}=\sigma_{\mathrm{IA}}$ for the WCA potential and $\varepsilon_{\mathrm{IIB}}=\varepsilon_{\mathrm{IA}}$ and $\sigma_{\mathrm{IIB}}=\sigma_{\mathrm{IA}}$ for the Lennard-Jones system. The goal of this calculation is to test whether it is possible to train the network to transform equilibrium configurations obtained via different potential energy functions. Indeed, even if the WCA interaction is in many aspect similar to the Lennard-Jones potential, the lack of an attractive region entails some differences in the behavior of systems in the fluid phase. For example, the WCA phase diagram does not have critical points and below a certain density admits only a single homogeneous fluid phase~\cite{khali_two_2021}.

The same observables computed in the previous example are also presented in this case using a sample of the same size. The potential energy distributions and the radial distribution functions are shown for the direct and inverse transformation in Fig.~\ref{fig:repulsive}.
These results show that the network can be trained to generate the correct energy and particle distributions of fluid systems also in the case of a different functional forms for the interaction potential between the source and target systems. The reweighted radial distribution function, in particular, shows that the structure of the liquid for the two different systems can be well reproduced by the normalizing flow.

In terms of efficiency of the process, once the training procedure is completed, the transformation between the two systems requires only a minimal computational overhead, mainly linked to the generation of the source configurations. Even though the energy of the target system still needs to be computed to obtain the reweighting factors (see Eqs.~\eqref{eq:omega_a} and~\eqref{eq:omega_b}), this has to be done only for independent configurations in the source space. In other words, the source space can be used to equilibrate and decorrelate configurations that are then transformed into the target space through the learned map. 
More in detail, transforming $2.5\times10^4$ configurations from system $\mathrm{IIA}$ to system $\mathrm{IIB}$ requires around 2\um{s} on a NVIDIA GeForce RTX 3080 Ti GPU, while the generation of the same number of target configurations (with the same degree of decorrelation) takes around 90\um{s} on the same hardware. Similar timings are observed in the first example.
The whole training protocol is summarized in the SM. For both numerical experiments shown in this paper, the training process took less than 6 hours.

In conclusion, the results presented in this letter show that normalizing flows can be used to generate independent configurations from statistical ensembles in the liquid state. 
This is made possible by the use of another physical system as a prior for the network, therefore encoding a high degree of physical information in the problem already before the training starts. It becomes then simpler and faster for the model to match the desired distribution after a relatively unexpensive training process and using a relatively small number of ground truth samples.
A successful reweighting procedure ensures that the averages computed over the generated configurations are free from any residual bias from the distribution of the source space. 
It is finally shown how the decorrelation of configurations in the target system can be achieved at the numerical cost of decorrelating samples in the source system.

This approach, presented in this letter via proof-of-principle calculations, paves the way to the exploitation of normalizing flows in transforming configurations obtained using computationally inexpensive models to more precise but computationally expensive descriptions. For example, a first step in this direction would be the transformation of configurations sampled through coarse-grained or all-atom force-field potential to \emph{first-principles} simulations, drastically reducing, in this way, the time needed for the numerical study of liquid systems at \emph{ab-initio} accuracy. In the spirit of Ref.~\citenum{wirnsberger_normalizing_2021} and of the calculations shown here, this could be achieved with a reduced set of ground samples (possibly empty) in such a way to minimize the need for \emph{ab-initio} evaluations of the energy. Note that, in future applications in this context, the difference between the probability distributions of the target and the source system is expected to be of the same order of the examples shown in this letter, as, for example, the force field parameters of the source system could be fine tuned to match as close as possible the distribution of the target space. This kind of numerical experiment will be the subject of future work.

\begin{acknowledgments}
We acknowledge financial support of the Austrian Science Fund (FWF) through the SFB TACO, Grant number F 81-N.
\end{acknowledgments}

%
%
%

\clearpage
\onecolumngrid
\normalsize
\patchcmd{\large}{15}{15}{}{}
\begin{center}
  \textbf{\LARGE Supplementary Material: Learning Mappings between Equilibrium States of Liquid Systems \\ Using Normalizing Flows}\\[.2cm]
  Alessandro Coretti,$^{1}$ Sebastian Falkner,$^{2}$ Phillip Geissler$^{3}$ and Christoph Dellago$^{1}$\\[.1cm]
  {\itshape ${}^1$University of Vienna, Faculty of Physics, 1090 Vienna, Austria.\\
  ${}^2$University of Vienna, Faculty of Physics \& Vienna Doctoral School in Physics, 1090 Vienna, Austria\\
  ${}^3$Department of Chemistry, University of California, Berkeley, California 94720, USA\\
  }
  ${}^*$Electronic address: alessandro.coretti@univie.ac.at\\
(Dated: \today)\\[2cm]
\end{center}

\setcounter{equation}{0}
\setcounter{figure}{0}
\setcounter{table}{0}
\setcounter{page}{1}
\setcounter{section}{0}
\renewcommand{\theequation}{S\arabic{equation}}
\renewcommand{\thefigure}{S\arabic{figure}}
\renewcommand{\thetable}{S\arabic{table}}
\renewcommand{\bibnumfmt}[1]{[S#1]}
\renewcommand{\citenumfont}[1]{S#1}
\renewcommand{\thesection}{S\Roman{section}}
\renewcommand{\thepage}{S\arabic{page}}

\titleformat*{\section}{\Large\bfseries}

\section{Data sets}
    \subsection{Simulation set-ups}
    The simulations presented in this work show how a normalizing flow can be trained to transform different systems of particles in the fluid state into each other. Here we consider 32 particles with unit mass in two dimensions. In particular, the two examples consist in I) the transformation between Lennard-Jones particles with unitary $\varepsilon_{\mathrm{IA}}$ and $\varepsilon_{\mathrm{IB}}=\varepsilon_{\mathrm{IA}}/2$ and II) the transformation between particles interacting via the WCA potential with a parameter of $\varepsilon_{\mathrm{IIA}}=\varepsilon_{\mathrm{IA}}/2$ and Lennard-Jones particles with $\varepsilon_{\mathrm{IIB}} = \varepsilon_{\mathrm{IA}}$. Both simulations consider the transformation and its inverse, so the systems are transformed into each other in both directions. In view of future applications, we here assume that in one case (A) it is very easy to sample from the Boltzmann distribution and in the other (B) it is not. We then assume that we have only a limited amount of data for the systems B while we can sample as much as we want from systems A. 
    
    In all cases, the systems are in the thermodynamic state specified by $\varrho^*=0.735$ and $T^*=1$ in the liquid region of the two-dimensional Lennard-Jones phase diagram~\cite{li_phase_2020_supp}. The asterisk indicates standard Lennard-Jones reduced units, where $\varepsilon$ and $\sigma$ of system $\mathrm{IA}$ are used to define units of energy and length, respectively, throughout the paper. A square box is employed in all the simulations which, at the given density, has an edge of size $L^*=6.6$ and it extends from 0 to $L^*$ in both dimensions. Periodic boundary conditions are enforced in all directions and a cutoff of $r^*_\mathrm{cut} = 3$ is employed for the Lennard-Jones interactions. The Lennard-Jones potential is also shifted by the value at the cutoff in order to avoid discontinuities in the potential energy function. No long-range corrections for the energy are used.
    
    \subsection{How are data obtained}
    Training data are obtained through standard Markov Chain Monte Carlo simulations. Each simulation is initialized using $2.5\times10^4$ walkers placed on a two-dimensional lattice and a simulation of $10^4$ Monte Carlo (MC) steps is performed to allow the systems to equilibrate. An MC step is here intended as performing a trial moves, i.e. as the attempt of displacing a particle in the two dimensions by a displacement $\Delta$. The parameter $\Delta$ is tuned for each system so that an acceptance ratio of about $0.5$ is achieved in the sampling process. The value of the parameter $\Delta$ for each system simulated is reported in Tab.~\ref{tab:deltas}.
    \begin{table}[htbp]
    \caption{Step sizes used in Markov chain Monte Carlo for the different simulations performed in this work. An acceptance ratio of around 0.5 is obtained for all the simulations.}
    \label{tab:deltas}
    \begin{tabular}{ccc}
    \hline
    System &  $\Delta$ & Acceptance\\
    \hline
    IA, IIB & 0.175 & 0.49\\
    IB      & 0.2   & 0.48  \\
    IIA     & 0.2   & 0.51 \\
    \hline
    \end{tabular}
    \end{table}
    The sampling process is started after equilibration and configurations are saved every $10^4$ MC steps in order to guarantee decorrelation between samples along the simulation. This is checked computing energy correlation functions of sampled configurations, from which it is possible to obtain the effective sample size which is around one for all the run performed. Training sets are composed by $9\times10^3$ configurations, and the network is validated using $10^3$ independent configurations. Once the training is complete, the transformations learned by the network are applied on different sets of configurations obtained with the same approach. Therefore, two sets of $5\times10^5$ configurations are sampled using an MCMC procedure like the one described above and the transformations $F$ and $F\inv$ are applied to move from system A to system B and viceversa, in both numerical simulations. Error bars for observables are estimated through a bootstrapping procedure where data are uniformly resampled with replacement 100 times. Reweighting is applied after bootstrapping. For the reweighting of radial distribution functions, relative weights are normalized by the average weight over the sample~\cite{van_constructing_2010_supp}.
    
\section{Transformation Layers}
\begin{figure}[htbp]
\centering
\includegraphics[width=\columnwidth]{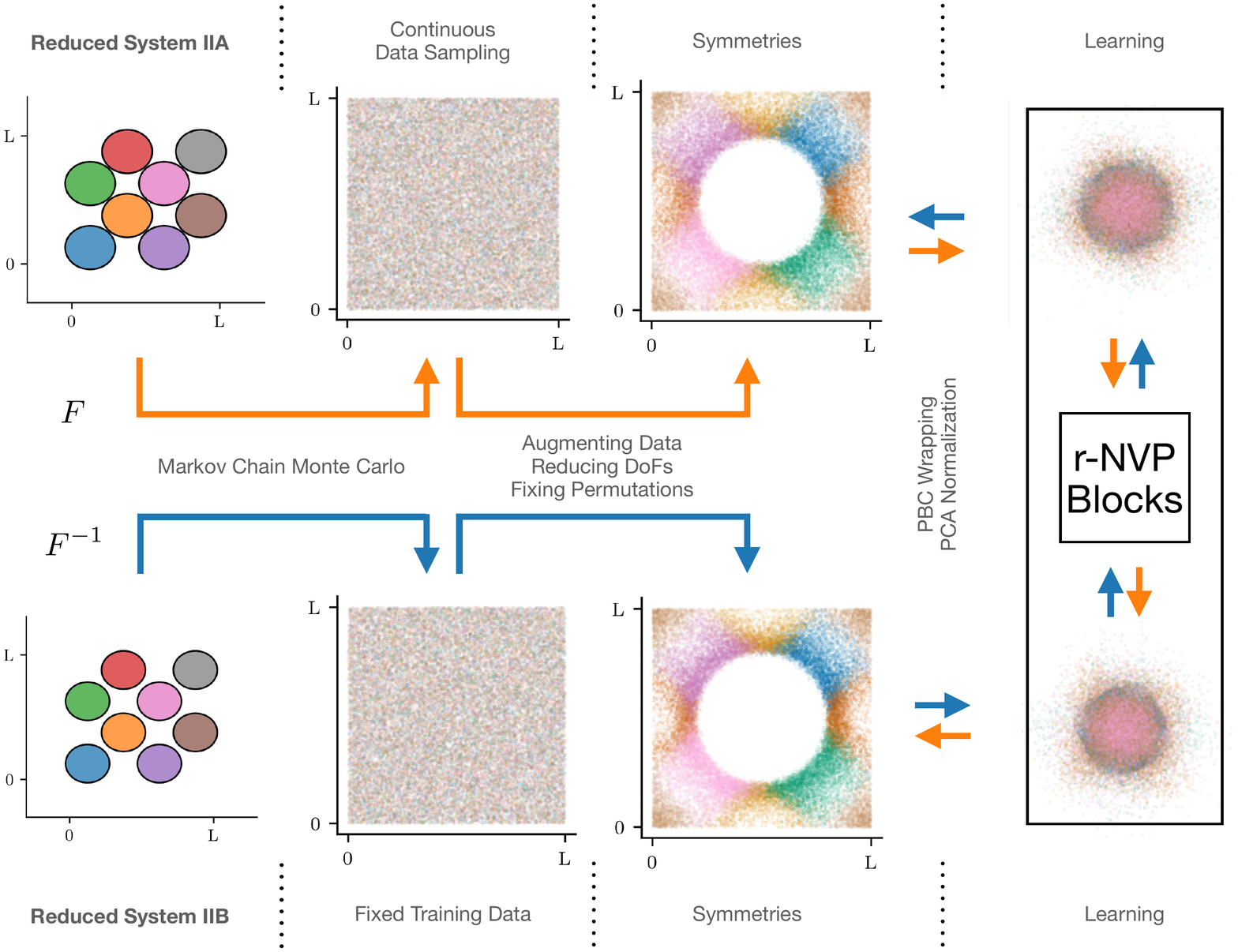}
\caption{
Schematic representation of the process that the input data undergo before being fed into the network for numerical experiment II. From left to right: initial configuration for WCA and Lennard-Jones particles. Attractive particles are represented in the picture as particles of smaller diameter (bottom) to distinguish them from the WCA, purely repulsive particles which are represented by a bigger size (top). The initial configurations are evolved using MCMC and, after equilibration, samples are produced (second column). For system IIA, configurations are continuously produced during training (top), while the training set is fixed for system IIB (bottom). Symmetries are then taken into account augmenting the training data by centering a random particle in the box and by applying one of the 8 transformations of the dihedral group of order eight. Degrees of freedom (DoFs) are reduced discarding the central particles and permutations are removed via Hungarian mapping~\cite{kuhn_hungarian_1955_supp}. Finally, data are normalized and fed into the real-NVP network. Normalization is performed in twp steps: PBCs are wrapped and PCA analysis is performed on the transformed input. Since these transformations are reversible, their inverse is also applied to the output, in order to obtain physically meaningful configurations out of the network.
}
\label{fig:transformations}
\end{figure}

Data are manipulated before feeding them into the normalizing flow. In particular, symmetries of the systems are either incorporated in the architecture of the network or they are removed from the data sets. This is known to improve drastically the efficiency of the training process, as the network is not very good in learning them from scratch~\cite{papamakarios_normalizing_2021_supp}. Augmentation layers are used to deal with discrete symmetries and to avoid overfitting, while normalization of the input also affects positively the training process. We describe these data manipulations in what follows. In Fig.~\ref{fig:transformations} a schematic representation of the transformations applied to the input is provided for a test system of 8 particles. This reduced number of degrees of freedom helps to visualize the transformations.

    \subsection{Symmetries: Translations and Rotations}
    The presence of PBCs limits the number of geometric symmetries that are satisfied by the system. In particular only translation invariance and a limited number of rotations and reflections leave the energy unchanged. Translations are taken into account shifting a random particle in the center of the box.
    Rotations and reflections in two dimensions are limited to the dihedral group of order eight (the octahedral group in three dimensions) which is composed by four reflections and four rotations for a total of eight transformations. 
    These two transformation layers are applied each time that a configurations is selected by the network from the corresponding set during training. In this way the training set is augmented by a factor of $32\cdot8=256$ as each configuration can be shifted to place each of the 32 particles at the center of the box and it can be rotated and/or reflected by one of the eight transformation of the dihedral group. No sign of overfitting is noted during the training process.
    It is also possible, at this point, to remove the two degrees of freedom relative to the particle in the origin to reduce the dimension of the space the network needs to learn.

    \subsection{Permutations}
    As particles are all of the same kind in our simulations, the energy is invariant with respect to relabeling of particles. Augmenting the training set using permutations as done for translations and rotations/reflections is unfeasible, due to the factorial nature of the permutation operation. Encoding this property directly in the architecture of the network is possible, but it has been shown to sometimes reduce the expressiveness of the model~\cite{kohler_equivariant_2020_supp}. For this reason, we follow a more brute-force approach which consists in removing permutations from the training set by relabeling particles in all the configurations by minimizing a ``distance'' with respect to a reference configuration (we chose the initial lattice as a reference). We included PBCs in the definition of the distance used by the algorithm. Permutation removal is realized through the use of the so-called Hungarian mapping~\cite{kuhn_hungarian_1955_supp}. In this way, all configurations in the sets are labeled in the same way with respect to their position in the box. One drawback is that the numerical complexity of the algorithm scales as the third power of the degrees of freedom. This nonlinear scaling with the system size does not represent a problem for the proof-of-principle calculations presented in this letter.
    
    \subsection{Normalization}
\begin{figure}[htbp]
    \centering
    \subfloat[][\emph{Effect of the PBC wrapping layer on all particles for all configurations. Since particles are all moved in the center of the box, unphysical configurations can appear. As the transformation is invertible, the output of the network can be unwrapped to obtain physical configurations.}]
    {\includegraphics[width=\textwidth]{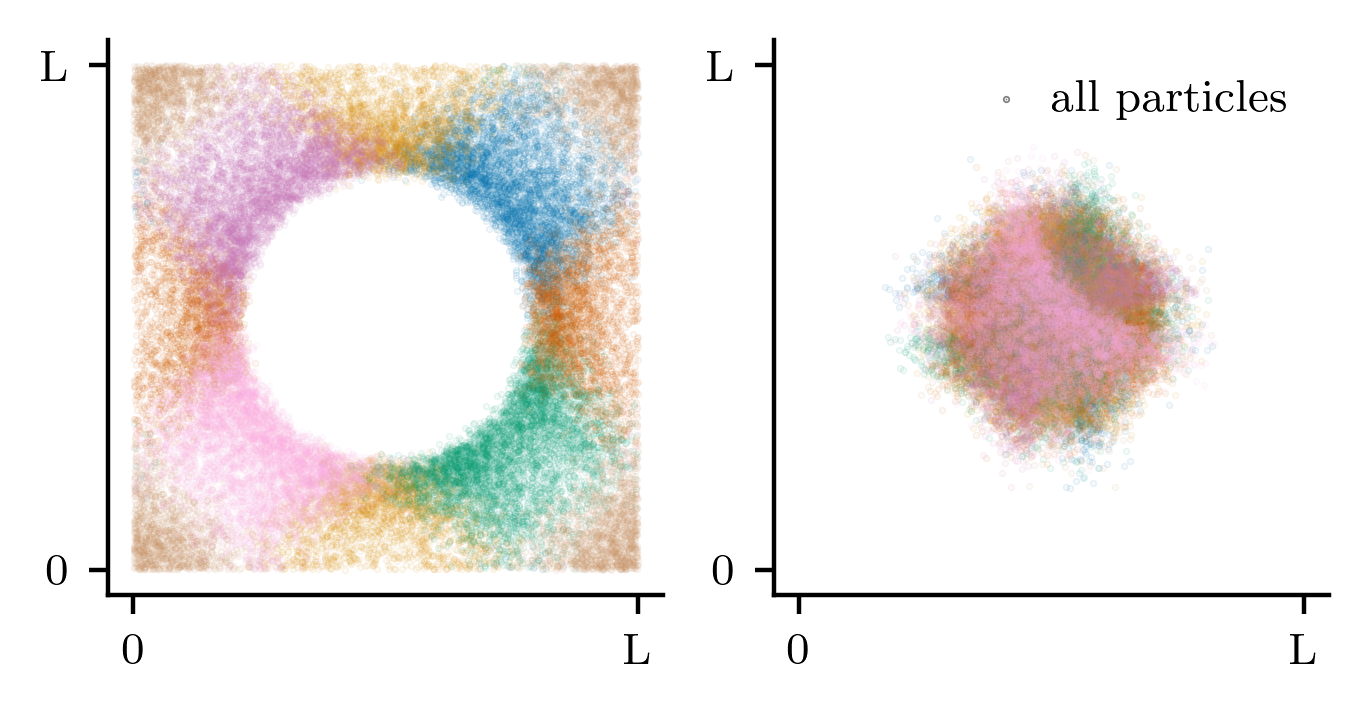}} \\
    \subfloat[][\emph{Effect of the PBC wrapping layer on a single particle index for all configurations of the sample. Note that in this particular case, configurations need to be wrapped on the $x$-axis but not on the $y$-axis. This choice is implemented on the basis of the spread of the samples in one direction. In other words, the transformation layer acts to minimize the variance between samples.}]
    {\includegraphics[width=\textwidth]{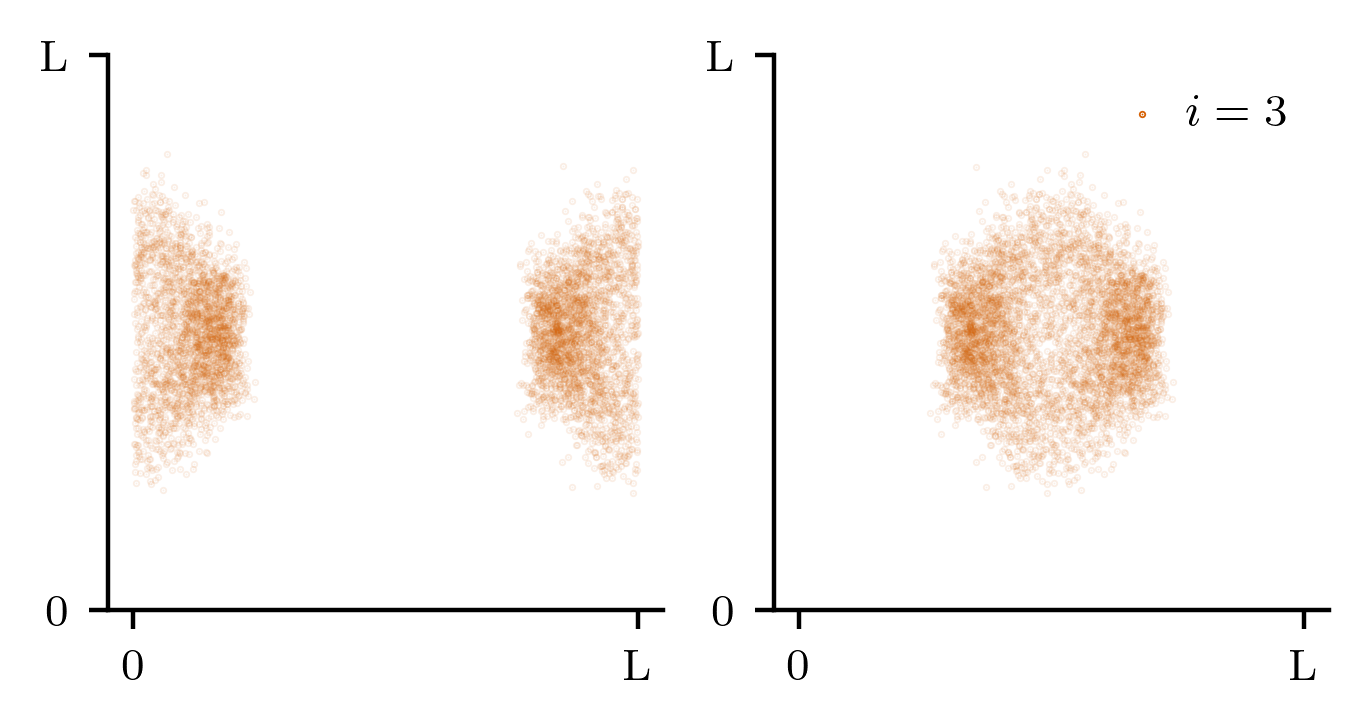}} \\
    \caption{Representation of the action of the PBC wrapping layer on a set of $5\times10^3$ configurations sampled from a system of 8 particles interacting via a WCA potential following the application of the Hungarian mapping.}
    \label{fig:PBCWrap}
    \end{figure}

    Normalization of the input is crucial in the framework of normalizing flows, as networks do not perform well when the range of values assumed by the input is not bounded~\cite{papamakarios_normalizing_2021_supp}. For this reason a principal component analysis (PCA) is usually recommended before feeding data into the network. The presence of PBCs, on the other hand, can produce complex patterns after PCA, which can result in even more complex inputs for the network to understand. For this reason, we include yet an additional transformation layer that takes the result of the Hungarian mapping and wraps the densities that have been separated by the PBC to focus it in the center of the box, minimizing in this way the spread of the samples for a given particle index. An example of the effect of this layer on configurations sampled from a system of 8 WCA particles is shown in Fig.~\ref{fig:PBCWrap}.
    From panel a) of Fig.~\ref{fig:PBCWrap} it is also possible to see how configurations, at this point, lose their physical nature. This is not an issue, as long as the wrapping procedure is reversible. The network learns to map wrapped configurations into wrapped configurations, which are then unwrapped to obtain physically meaningful samples of the target system.
    Once the wrapping process is completed, a PCA can be performed on the transformed input and data can finally be fed into the network. For the PCA analysis we use the configurations of the source and target systems as a reference to compute the principal components of the distribution as we noticed a quite remarkable improvement of training performances in this way.
    
\section{Network}

\subsection{Architecture}
\begin{figure}[htbp]
\centering
\includegraphics[width=\columnwidth]{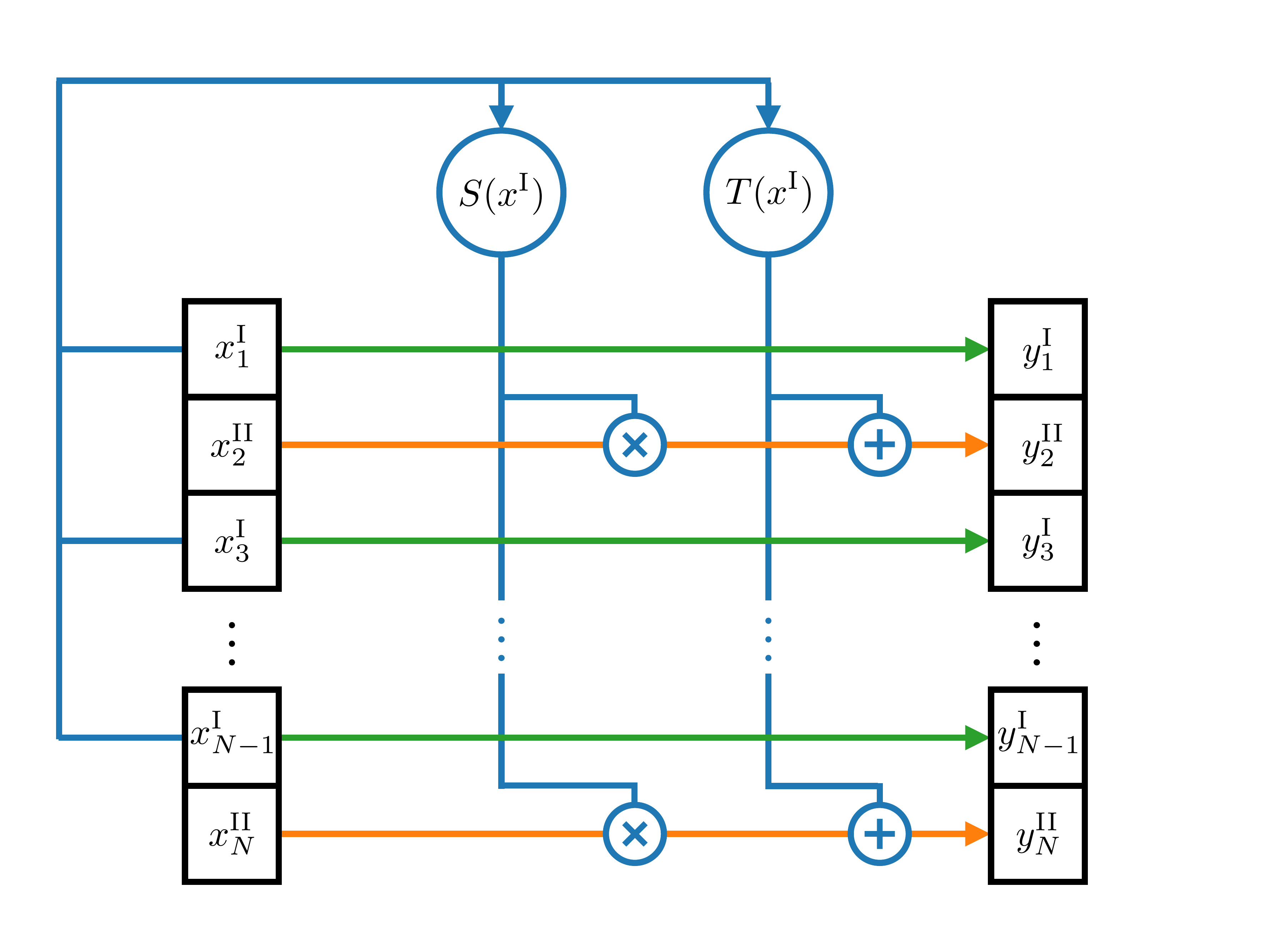}
\caption{Schematic representation of the first real-NVP block of the network architecture. The input is masked alternately, so that odd entries $x_1, x_3, \dots$ are left unchanged and used as arguments for the functions $S$ and $T$. These functions are then used to modify the even entries of the input $x_2, x_4, \dots$ through simple operations of scaling and translation. The masking process is represented by the superscripts $\mathrm{I}$ (leave unchanged and use as input) and $\mathrm{II}$ (apply transformation) on the input ($x$) and output ($y$) of the block. The functions $S$ and $T$ are represented by feed-forward networks whose parameters are the object of the optimization during the training process. Multiple blocks are stacked together so to act on different entries of the input alternately. For example, the second block leaves the even entries of the input unchanged and uses them as input for the functions $S$ and $T$ that are then used to modify the odd input entries and so on.}
\label{fig:rnvp-block}
\end{figure}

The network used in this work is based on the real-NVP architecture introduced in Ref.~\citenum{dinh_density_2017_supp} and it is the same for both examples presented in the main text. 8 real-NVP blocks are stacked together and the input is masked alternately, so that odd blocks leave odd components of the input unchanged. In Fig.~\ref{fig:rnvp-block} a schematic representation of the first block is shown. 
Two separate tanh-activated feed-forward networks are used to model the nonlinear functions $S$ and $T$ in each block. Each feed-forward network is composed by 3 hidden layer with 128 nodes per layer. This yields a total of 457,696 parameters to be optimized during the training process. The optimization is performed via the Adam algorithm~\cite{kingma_adam_2014_supp}.
\subsection{Training}
\begin{figure}[htbp]
\centering
\subfloat[][\emph{Case I}]
{\includegraphics[width=\textwidth]{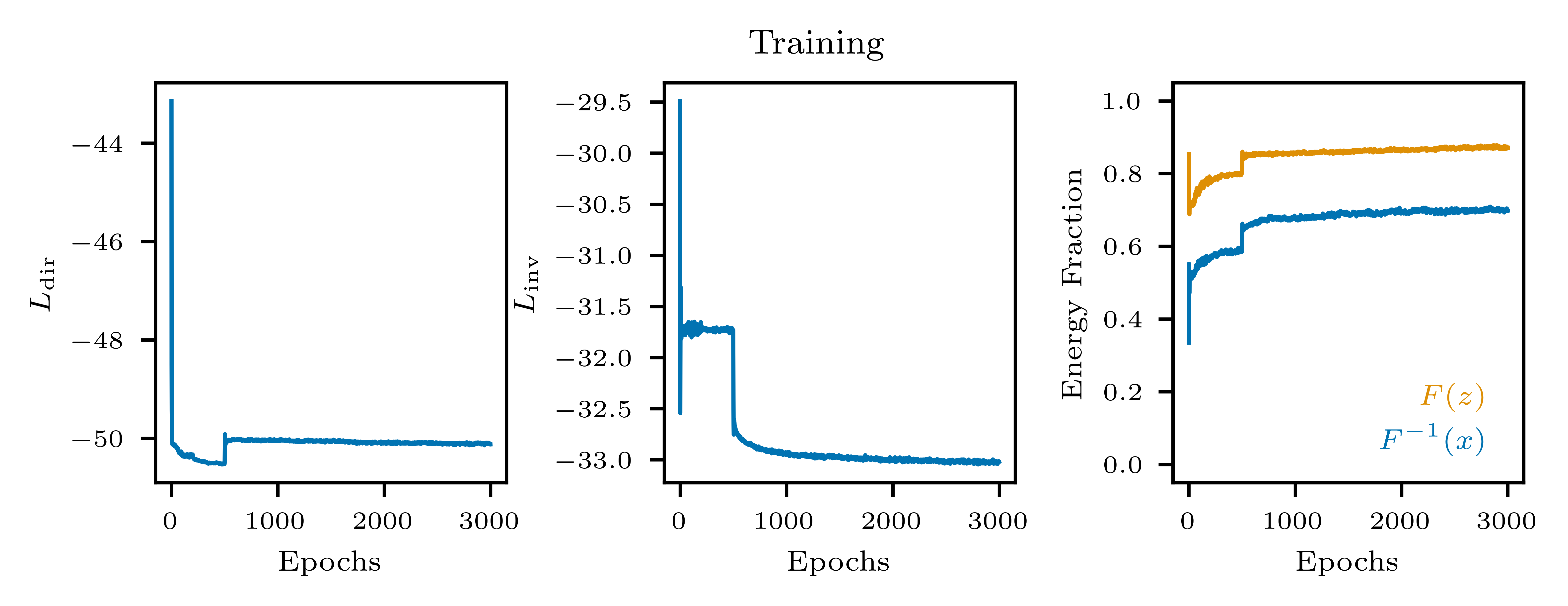}} \\
\subfloat[][\emph{Case II}]
{\includegraphics[width=\textwidth]{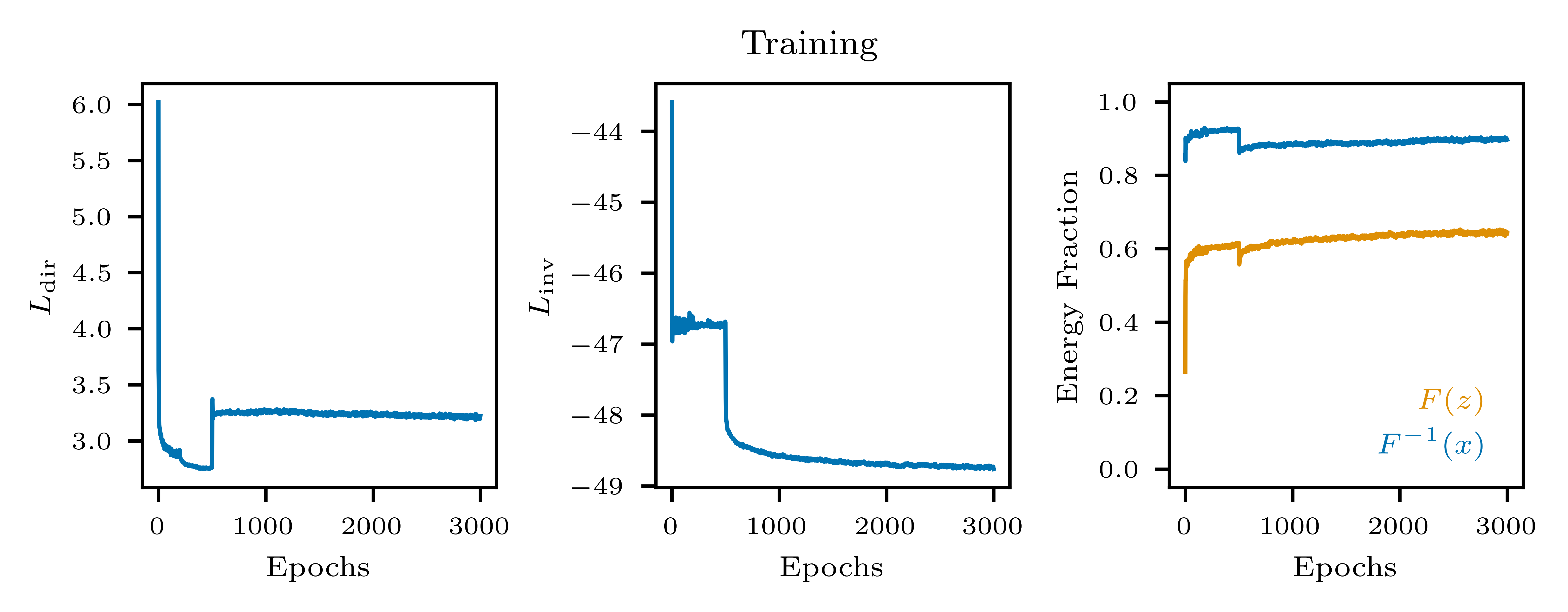}} \\
\caption{Training metrics for the numerical experiments presented in the main text. The first and second column show the trend of the two terms of the loss function. Third column represents the fraction of generated data within the target energy range. A point is assumed to be within the target energy range if it is within the 1st and 99th energy percentile of separately sampled reference data.}
\label{fig:training}
\end{figure}

The training protocol is also the same for both numerical experiments. For the whole training, we use a batch size of 2048 and a value for the clamping start of the potential energy of $10^5$. To additionally prevent instabilities due to particle overlaps during training, the potential energy is also linearized for $r^*<0.8$ enforcing continuity of the first derivative at the inner cutoff $r^*_\text{incut} = 0.8$. The training protocol consists in a total of 3000 epochs divided in two parts. The first part consists in 500 epochs with $\lambda_{\text{dir}} = 1$ and $\lambda_{\text{inv}} = 0$. The learning rate is kept fixed to the value of $10^{-3}$ for the first 200 epochs and it is then lowered to $10^{-4}$ for the remaining epochs. The second part of the training consists in 2500 epochs with $\lambda_{\text{dir}} = 1$ and $\lambda_{\text{inv}} = 1$ with a fixed learning rate of $10^{-4}$. The training metrics are summarized in Fig.~\ref{fig:training}.
During the second part of the training, as samples from the system $\mathrm{A}$ are needed to compute the loss function $L_{\text{inv}}$, the continuous sampling process introduced in the main text is used. Therefore configurations are continuously produced using an MCMC approach. The sampling procedure is started from a random batch of initial states selected from a pool of samples. New configurations are saved every $10^3$ Monte Carlo steps. In the pool, initial configurations are replaced by the newly sampled ones. For this reason, the initial pool needs not to be vast nor initial states need to be really independent. If the number of elements in the pool is lower than the batch size, resampling with replacement can be used. Moreover, as the training process proceeds, configurations will become independent anyway.

\end{document}